\documentclass[sigconf]{acmart}

\usepackage{enumitem}
\usepackage{multirow}
\usepackage{graphicx}
\usepackage{float}
\usepackage{subcaption}

\AtBeginDocument{%
  \providecommand\BibTeX{{%
    \normalfont B\kern-0.5em{\scshape i\kern-0.25em b}\kern-0.8em\TeX}}}

\copyrightyear{2023} 
\acmYear{2023} 
\setcopyright{acmlicensed}\acmConference[CIKM '23]{Proceedings of the 32nd ACM International Conference on Information and Knowledge Management}{October 21--25, 2023}{Birmingham, United Kingdom}
\acmBooktitle{Proceedings of the 32nd ACM International Conference on Information and Knowledge Management (CIKM '23), October 21--25, 2023, Birmingham, United Kingdom}
\acmPrice{15.00}
\acmDOI{10.1145/3583780.3614781}
\acmISBN{979-8-4007-0124-5/23/10}

\begin{document}

\title[APGL4SR]{APGL4SR: A Generic Framework with Adaptive and Personalized Global Collaborative Information in Sequential Recommendation}

\author{Mingjia Yin}
\email{mingjia-yin@mail.ustc.edu.cn}
\orcid{0009-0005-0853-1089}
\affiliation{%
  \institution{Anhui Province Key Laboratory of
Big Data Analysis and Application, University of Science and Technology of China \& State Key Laboratory of Cognitive Intelligence}
  \city{Hefei}
  \country{China}
}

\author{Hao Wang}
\authornote{Corresponding author.}
\email{wanghao3@ustc.edu.cn}
\orcid{0000-0001-9921-2078}
\affiliation{%
  \institution{Anhui Province Key Laboratory of
Big Data Analysis and Application, University of Science and Technology of China \& State Key Laboratory of Cognitive Intelligence}
  \city{Hefei}
  \country{China}
}

\author{Xiang Xu}
\email{demon@mail.ustc.edu.cn}
\orcid{0009-0008-5508-3145}
\affiliation{%
  \institution{Anhui Province Key Laboratory of
Big Data Analysis and Application, University of Science and Technology of China \& State Key Laboratory of Cognitive Intelligence}
  \city{Hefei}
  \country{China}
}

\author{Likang Wu}
\email{wulk@mail.ustc.edu.cn}
\orcid{0000-0002-4929-8587}
\affiliation{%
  \institution{Anhui Province Key Laboratory of
Big Data Analysis and Application, University of Science and Technology of China \& State Key Laboratory of Cognitive Intelligence}
  \city{Hefei}
  \country{China}
}

\author{Sirui Zhao}
\email{sirui@mail.ustc.edu.cn}
\orcid{0000-0001-8103-0321}
\affiliation{%
  \institution{Anhui Province Key Laboratory of
Big Data Analysis and Application, University of Science and Technology of China \& State Key Laboratory of Cognitive Intelligence}
  \city{Hefei}
  \country{China}
}

\author{Wei Guo}
\email{guowei67@huawei.com}
\orcid{0000-0001-8616-0221}
\author{Yong Liu}
\email{liu.yong6@huawei.com}
\orcid{0000-0001-9031-9696}
\affiliation{%
  \institution{Huawei Singapore Research Center}
  \country{Singapore}
}


\author{Ruiming Tang}
\email{tangruiming@huawei.com}
\orcid{0000-0002-9224-2431}
\affiliation{%
  \institution{Huawei Noah's Ark Lab}
  \city{Shenzhen}
  \country{China}
}

\author{Defu Lian}
\email{liandefu@ustc.edu.cn}
\orcid{0000-0002-3507-9607}
\affiliation{%
  \institution{Anhui Province Key Laboratory of
Big Data Analysis and Application, University of Science and Technology of China \& State Key Laboratory of Cognitive Intelligence}
  \city{Hefei}
  \country{China}
}

\author{Enhong Chen}
\email{cheneh@ustc.edu.cn}
\orcid{0000-0002-4835-4102}
\affiliation{%
  \institution{Anhui Province Key Laboratory of
Big Data Analysis and Application, University of Science and Technology of China \& State Key Laboratory of Cognitive Intelligence}
  \city{Hefei}
  \country{China}
}


\renewcommand{\shortauthors}{Mingjia Yin, et al.}
\begin{abstract}
The sequential recommendation system has been widely studied for its promising effectiveness in capturing dynamic preferences buried in users' sequential behaviors. Despite the considerable achievements, existing methods usually focus on intra-sequence modeling while overlooking exploiting global collaborative information by inter-sequence modeling, resulting in inferior recommendation performance. Therefore, previous works attempt to tackle this problem with a global collaborative item graph constructed by pre-defined rules. However, these methods neglect two crucial properties when capturing global collaborative information, i.e., adaptiveness and personalization, yielding sub-optimal user representations. To this end, we propose a graph-driven framework, named \textbf{A}daptive and \textbf{P}ersonalized \textbf{G}raph \textbf{L}earning for \textbf{S}equential \textbf{R}ecommendation (APGL4SR), that incorporates adaptive and personalized global collaborative information into sequential recommendation systems. Specifically, we first learn an adaptive global graph among all items and capture global collaborative information with it in a self-supervised fashion, whose computational burden can be further alleviated by the proposed SVD-based accelerator. Furthermore, based on the graph, we propose to extract and utilize personalized item correlations in the form of relative positional encoding, which is a highly compatible manner of personalizing the utilization of global collaborative information. Finally, the entire framework is optimized in a multi-task learning paradigm, thus each part of APGL4SR can be mutually reinforced. As a generic framework, APGL4SR can not only outperform other baselines with significant margins, but also exhibit promising versatility, the ability to learn a meaningful global collaborative graph, and the ability to alleviate the dimensional collapse issue of item embeddings. The code is available at https://github.com/Graph-Team/APGL4SR.
\vspace{-0.5cm}
\end{abstract}

\ccsdesc[500]{Information systems~Personalization}

\keywords{Recommendation System; Sequential Recommendation; Graph Contrastive Learning}


\maketitle

\section{INTRODUCTION}
The recommendation system has been an important research direction, aiming to model users' preferences from their historical interaction records and further recommend personalized items to them\cite{RS_survey, RS_GNN_survey, wlk_survey}. However, in a real-world scenario, users' preferences exhibit a trend of dynamic changes over time, rendering it hard to obtain accurate user preference representations. In this context, the sequential recommendation (SR) system has been widely studied for its promising effectiveness in capturing dynamic preferences buried in users' sequential behaviors\cite{sequential_recommendation_survey1, sequential_recommendation_survey2, sequential_recommendation_survey3, Caser, GRU4Rec, SASRec}.

\begin{figure}
    \centering
    \includegraphics[scale=0.35]{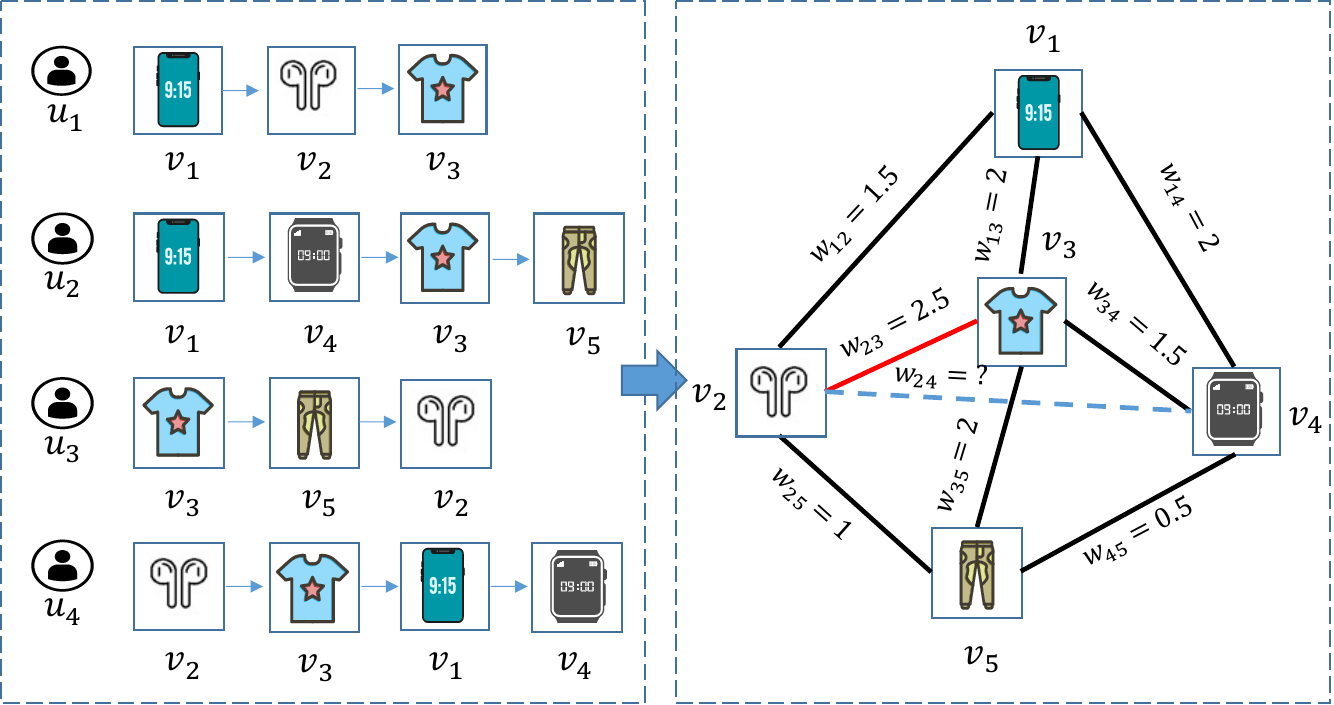}
    \caption{A toy example to illustrate the importance of global collaborative modeling in sequential recommendation systems. The graph is constructed based on the rules in Sec. \ref{sec: initial graph construction}.}
    \label{fig: motivation}
    \vspace{-0.5cm}
\end{figure}

Generally speaking, the crux of sequential recommendation lies in fully discovering the item transition patterns within users' interaction sequences, and encoding them into user representations. Recently, with the development of deep learning, enormous amounts of work are devoted to learning meaningful user representations with various deep neural networks, including Convolutional Neural Network\cite{Caser, 3DCNN, CosRec}, Recurrent Neural Network\cite{GRU4Rec, narm, STAMP}, Transformer\cite{SASRec, bert4rec, TiSASRec} and Graph Neural Network (GNN)\cite{SRGNN, GCSAN, LESSER}. Among these works, the Transformer-based methods have been dominant for their promising representational power, GNN-based methods have also been widely studied for their capability to capture complex item transition patterns within interaction sequences.

Despite the promising results achieved by the aforementioned methods, most of them only focus on mining information from an individual sequence. However, other sequences with similar item transition patterns, i.e., sub-sequences, can also contribute to the recommendation of the current sequence. Overlooking such information, which we define as \textbf{global collaborative information}, will limit the representational power of the learned user representations. For instance, Fig. \ref{fig: motivation} is a sequential recommendation scenario, wherein four users $u_1$, $u_2$, $u_3$, and $u_4$ interacted with some items chronologically. The purpose of an SR model is to predict one next item for each user according to their historical interaction records. Based on pre-defined rules like connecting items with co-occurrence relationships, we can construct a global item graph with the sequences. In the graph, we can find the T-shirt is relevant to pants. When recommending a new item to user $u_1$, who just clicked a T-shirt, the RS can increase the confidence of recommending a pair of pants to $u_1$. However, we argue that the global collaborative information obtained from the rule-based graph is sub-optimal due to the absence of two essential properties: (1) \textbf{Adaptiveness}. The graph should be able to adaptively detect noisy and underlying edges, e.g., the edge weight between the T-shirt and the earphones could be decreased as they are not relevant intuitively, and we may add an edge between the intuitively similar earphones and the smartwatch. (2) \textbf{Personalization}. The global collaborative information should have varying effects on distinct users, e.g., a fan of electronics like user $u_4$ may insist on clicking another electronic even though he or she just clicked a T-shirt. Motivated by these issues, we intend to incorporate adaptive and personalized global collaborative information into sequential recommendation systems.

Nevertheless, how to obtain adaptive global collaborative information in the form of a graph and meticulously utilize it remains to be further explored. Several attempts have been made to address the above issues\cite{GCEGNN, GCL4SR, DGNN}, which we argue is not the optimal solution. Considering the graph acquisition, both GCE-GNN\cite{GCEGNN} and GCL4SR\cite{GCL4SR} construct a fixed global item transition graph with pre-defined rules, which may include noisy edges or neglect underlying edges as mentioned above. DGNN\cite{DGNN} proposes to learn explicit and implicit connections among global items, which can increase its adaptiveness. However, its modeling scope is limited to items within a single batch for efficiency rather than all items, thereby rendering it challenging to capture diverse collaborative information among items. As for graph utilization, only a few works have considered extracting personalized information from the global graph, e.g., GCL4SR generates user-specific graph representations and directly combines them with sequential representations. However, we observe that the combination-based strategy will yield inferior results. Intuitively, the two representations lie in different latent spaces, simply combining them with linear projection may lead to information degeneration.

Therefore, to incorporate global collaborative information into sequential recommendation systems, we are faced with two main challenges: (1) How to capture adaptive global collaborative information effectively and efficiently. The rule-based graph construction will lead to biased representations, while directly modeling the implicit connections between all pairs of items will result in unacceptable time and space complexity. (2) How to extract and utilize personalized information from the global collaborative information. As mentioned before, to avoid the information degeneration issue, we need to realize a personalized utilization of global collaborative information in a more compatible way.

In order to fully mine the global collaborative information and address the above challenges, we propose a graph-driven framework, named \textbf{A}daptive and \textbf{P}ersonalized \textbf{G}raph \textbf{L}earning for \textbf{S}equential \textbf{R}ecommendation (APGL4SR), that incorporates adaptive and personalized global collaborative information into sequential recommendation systems with self-supervised learning. Firstly, we propose an adaptive global collaborative learner, which adaptively learns a refined global graph among all items and captures global collaborative information via mutual information maximization, and then the computational complexity can be reduced with the proposed SVD-based acceleration strategy, which in total incorporate global collaborative information into item representations effectively and efficiently. Then, we present a personalized graph extractor, extracting item correlations from the global collaborative graph in the form of relative positional encoding and transforming them with user-specific embedding. By injecting obtained relative position encoding into the Transformer encoder, we can realize personalized utilization of the global collaborative information for each user in a highly compatible manner. Finally, we leverage a multi-task learning paradigm to optimize the entire framework in an end-to-end fashion, by which the adaptive global collaborative learner and the subsequent sequential model can be mutually reinforced. Our main contributions are summarized as follows:
\begin{itemize}[leftmargin=*]
    \item We study a novel problem in the sequential recommendation to learn adaptive and personalized global collaborative information.
    \item We develop an adaptive global collaborative learner to capture global collaborative information. With the aid of the module, we can learn a global collaborative graph carrying meaningful information by adaptively refining the graph, whose efficiency is guaranteed by the proposed SVD-based accelerator.
    \item We propose a personalized graph extractor to meticulously utilize global collaborative information for each specific user. With the proposed module, we can extract and leverage personalized information from the global graph in a highly compatible manner.
    \item As a generic framework, APGL4SR can not only outperform other baselines with significant margins on four public datasets, but also exhibit promising versatility, the ability to learn a meaningful global collaborative graph, and the ability to alleviate the dimensional collapse issue of item embeddings.
\end{itemize}

\section{RELATED WORKS}
\subsection{Sequential Recommendation System}\label{sec: related RS}
As a main branch of recommendation systems, sequential recommendation (SR) aims to capture dynamic preferences in chronologically organized user behaviors for more accurate and timely results\cite{sequential_recommendation_survey1, sequential_recommendation_survey2, sequential_recommendation_survey3, wu_influence, lian2020personalized}. 
Before the dawn of deep learning, conventional methods attempted to employ machine learning approaches to model item transition patterns in sequences, such as KNN-based methods\cite{itemknn, sknn, stan} and Markov-Chain-based methods\cite{FPMC, Fossil}.

With the development of deep learning, some deep sequential recommendation models were presented to model the complex sequential interest with deep neural networks: Caser\cite{Caser} and CosRec\cite{CosRec} treated sequences as images and adopted Convolutional Neural Network (CNN) to process the sequences, GRU4Rec\cite{GRU4Rec} and NARM\cite{narm} instead considered Recurrent Neural Network (RNN) and its variants to capture dynamic user preferences. Among all the deep sequential recommendation system models, Transformer\cite{transformer}-based approaches like SASRec\cite{SASRec} and BERT4Rec\cite{bert4rec} are currently dominant for their powerful modeling capability. Besides, Graph Neural Networks (GNN) have also been prevalent for their capability to capture high-order structures\cite{GNN_survey, RS_GNN_survey, MCNE}, which are widely adopted to model complex item transition patterns in users' sequences. For example, SR-GNN\cite{SRGNN} transformed each user sequence into a directed graph with pre-defined rules and then utilized multiple GatedGNN layers upon the graphs, enabling it to capture more complicated item transition patterns compared with RNN-based methods. GC-SAN\cite{GCSAN} enhanced the item representations of SR-GNN with stacked self-attention layers. LESSER\cite{LESSER} further proposed two kinds of session graphs to alleviate the information loss and long-range dependency capturing problems. 

Despite the success, these methods all focused on intra-sequence modeling while neglecting inter-sequence information, limiting their representation ability. In this context, GCE-GNN\cite{GCEGNN} introduced an extra rule-based global graph to model local and global item relationships simultaneously, while DGNN\cite{DGNN} adopted a graph neural network with a single gate and an adaptive graph neural network to capture explicit and implicit dependencies among all items within a single batch. Unlike both of them, GSL4Rec\cite{GSL4Rec} focused on session-based social recommendation and learned a global user-user graph to incorporate global information. In contrast, we propose to capture the global collaborative information among all items in a self-supervised learning fashion.

\subsection{Self-Supervised Learning in SR}\label{sec: related SSL in SR}
The Self-Supervised Learning (SSL) paradigm has been widely studied in multiple research communities, e.g., Computer Vision\cite{CPC, CPCv2, DIM, MoCo, SimCLR, SimSiam}, Natural Language Processing\cite{bert, BART}, and non-sequential recommendation systems\cite{ssl_rec_survey, SGL, SimGCL, SEPT, CCDR, LightGCL}. SSL has recently been introduced into sequential recommendation systems to alleviate the data sparsity issue. Some pioneer works construct self-supervised tasks at the data level. S$^3$-Rec\cite{S3Rec} devised four auxiliary pre-training tasks to model correlations among different views of raw data. Unlike S$^3$-Rec, CL4SRec\cite{CL4SRec} focused on exploring contrastive learning on user behavior sequences with sequence level augmentations, e.g., crop, mask, or reorder, to yield better user representations. Except for data level SSL tasks, ICLRec\cite{ICLRec} performed contrastive learning by contrasting user sequence representations and the intent prototype representations, DuoRec\cite{DuoRec} constructed augmented representations efficiently by treating Transformer output with different dropout masks as positive samples, GCL4SR\cite{GCL4SR} and GUESR\cite{GUESR} leveraged graph contrastive learning on rule-based global item graphs to capture global collaborative information. Instead, our approach is to develop an adaptive global collaborative learner to obtain better global collaborative information and personalize the utilization of global collaborative information effectively.

\subsection{Graph Structure Learning}\label{sec: GSL}
In this paper, rather than rule-based graphs introduced in Sec. \ref{sec: related RS}, we intend to learn an adaptive global item graph to capture global collaborative information, so we introduce the graph structure learning technique to achieve this. Graph Structure Learning (GSL) has been studied recently in pursuit of an optimal graph structure\cite{gsl_survey}. There are two critical steps to obtain an optimal graph: graph structure modeling and post-processing. The graph structure modeling aims to model the weight of all edges based on metric-based approaches, neural-based approaches, or direct approaches. Metric-based methods leverage different kernel functions like Gaussian kernels\cite{AGCN} or inner-product kernels\cite{GAUG_M} to measure node similarities. Neural-based methods incorporate complex deep neural networks to construct the intrinsic connection of nodes, e.g., GAT\cite{GAT} or Graphormer\cite{Graphormer}. Rather than learning pair-wise node relationships, direct approaches\cite{GLNN, ProGNN, LRGNN} regard the adjacency matrix as free variables to learn with some structural hypothesis on the learned graph. After obtaining an intermediate graph, the graph will be post-processed to ensure desired properties. General post-processing techniques include discrete sampling and residual connection. In this paper, we introduce GSL to address a novel problem, where we aim to learn an adaptive and personalized global item graph in the sequential recommendation to capture refined global collaborative information. 

\section{PRELIMINARIES}
\subsection{Problem Definition} \label{sec: next item prediction}
In sequential recommendation, the main task is to infer user preference representations from users' historical interaction records and recommend the next item to them. Specifically, considering a sequential recommendation system, we denote $\mathcal{U}$ as a set of users and $\mathcal{V}$ as a set of items, and $|\mathcal{U}|$ and $|\mathcal{V}|$ is the number of users and items respectively. As users' interaction records can be organized chronologically, we can define the interaction sequence of a user $u \in \mathcal{U}$ as $s_u = [v_1, v_2, \dots, v_t, \dots, v_{|s_u|}]$, where $v_t \in \mathcal{V}$ is the item interacted by user $u$ at time step $t$, $|s_u|$ is the length of the sequence and has a maximum length of $N$. Then we can define the next item prediction task, which aims to predict the item at the next time step $v_{|s_u| + 1}$ for each user $u \in \mathcal{U}$.

\subsection{Rule-Based Global Item Transition Graph}\label{sec: initial graph construction}
Conventional SR methods focus on intra-sequence modeling while overlooking exploiting global collaborative information modeling by inter-sequence modeling, resulting in inferior user representations. To this end, we propose to capture the global collaborative information with a global item transition graph. However, we are only given the interaction records of users without an available global item graph. Therefore, we intend to learn the graph with GSL introduced in Sec. \ref{sec: GSL}, where the noisy edges and underlying edges mentioned in Fig. \ref{fig: motivation} can be adaptively detected.

As we intend to obtain an ideal global item transition graph with GSL, it is essential to construct an initial graph with relatively adequate prior knowledge for stable training. Therefore, we leverage the rule-based graph which is defined and proved useful in GCL4SR\cite{GCL4SR} as our primary graph. Then we will elaborate on how to construct the graph $\mathcal{G}$ with its adjacent matrix $\mathcal{A} \in \mathbb{R}^{|\mathcal{V}| \times |\mathcal{V}|}$, which is initialized as an all-zero matrix.

As illustrated in the right part of Fig. \ref{fig: motivation}, considering the user sequence $s_u = [v_1, v_2, \dots, v_t, \dots, v_{|s_u|}]$ defined in Sec. \ref{sec: next item prediction}, we update the edge weight as $\mathcal{A}_{v_1, v_k} \leftarrow \mathcal{A}_{v_1, v_k} + \frac{1}{k}$ for each item pair $(v_1, v_k)$ within a sliding window, where $k$ is the size of the sliding window and we set it to 2 empirically. Taking the sequence of user $u_1$ in Fig. \ref{fig: motivation} as an example, for $(v_1, v_2)$, we update $\mathcal{A}_{v_1, v_2} \leftarrow \mathcal{A}_{v_1, v_2} + 1$, and for $(v_1, v_3)$, we perform the operation $\mathcal{A}_{v_1, v_3} \leftarrow \mathcal{A}_{v_1, v_3} + \frac{1}{2}$. Similarly, we can repeat the above operation for each user sequence to build an intermediate global item graph, and then we normalize the adjacency matrix as follows,
\begin{equation}\label{eq: graph normalization}
    \mathcal{A}_{v_i, v_j} \leftarrow (\frac{1}{deg(v_i)} + \frac{1}{deg(v_j)}) \mathcal{A}_{v_i, v_j},
\end{equation}
where $deg(\cdot)$ is the degree of some node in the graph $\mathcal{A}$. Finally, we make the graph $\mathcal{G}$ symmetric and add a self-loop for each node.

\subsection{LightGCN}\label{sec: LightGCN}
As we intend to capture global collaborative information in the form of graphs, we need a graph encoder to generate graph representations. Specifically, we select LightGCN\cite{LightGCN} for its effectiveness. In detail, given the $(l - 1)$-th layer node embedding $\mathbf{E}^{(l-1)}$, the message-passing process of LightGCN is defined as follows,
\begin{equation}\label{eq: LightGCN message passing}
    \mathbf{E}^{(l)} = \mathcal{A} \mathbf{E}^{(l - 1)}
\end{equation}
where $\mathcal{A} \in \mathbb{R}^{|\mathcal{V}| \times |\mathcal{V}|}$ is the adjacency matrix. After multiple layers of message-passing, the final graph representation is obtained by:
\begin{equation}\label{eq: LightGCN final representation}
    \mathbf{E}^{(L)} \leftarrow \frac{1}{L} (\mathbf{E}^{(0)} + \mathbf{E}^{(1)} + \dots + \mathbf{E}^{(L)})
\end{equation}
where $L$ is the number of LightGCN layers.

\section{METHODOLOGY}
\begin{figure*}
    \centering
    \includegraphics[scale=0.4]{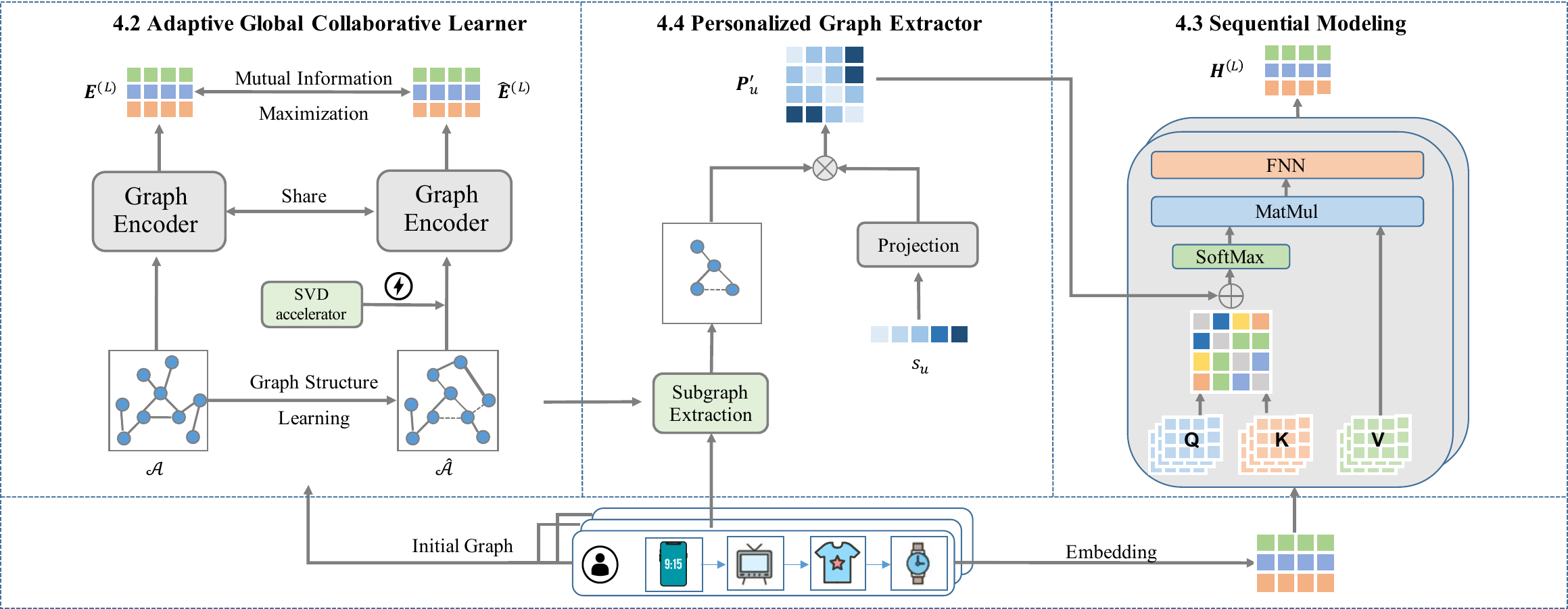}
    \caption{The framework of the proposed APGL4SR model.}
    \label{fig: framework}
    \vspace{-0.5cm}
\end{figure*}
In this section, we first present a general description of the proposed model. Then we introduce each part of the model in turn and finally illustrate the optimization of the entire framework.

\subsection{Overall Framework}
In this paper, we propose the \textbf{A}daptive and \textbf{P}ersonalized \textbf{G}raph \textbf{L}earning for \textbf{S}equential \textbf{R}ecommendation (APGL4SR) framework to incorporate adaptive and personalized global collaborative information into sequential recommendation systems with SSL, which is illustrated in Fig. \ref{fig: framework}. Specifically, we first propose a novel adaptive global collaborative learner (AGCL) in Sec. \ref{sec: global collaborative learner}. In the AGCL, we utilize the GSL technique introduced in Sec. \ref{sec: GSL} to learn a refined global item correlation graph, whose computational complexity can be further reduced by the proposed SVD-based acceleration strategy. Given the original graph and the refined graph, we can generate two representations respectively, then we maximize their mutual information to incorporate the adaptive global collaborative information into item representations. With the enhanced item representations, we utilize a classic Transformer encoder in Sec. \ref{sec: sequential modeling} to generate user representations. However, distinct users may exhibit different interests in global collaborative information, so we further propose a personalized graph extractor (PGE) in Sec. \ref{sec: graph adaption}. The extractor extracts personalized information from the refined global graph for individual users in the form of relative positional encoding, thus it can be naturally combined with the sequential encoder in Sec. \ref{sec: sequential modeling}. Finally, we elaborate on how to optimize the entire framework via a multi-task learning paradigm in Sec. \ref{sec: multi-task learning}. In this way, the adaptive global collaborative learner and the sequential model can be mutually reinforced. 

\subsection{Adaptive Global Collaborative Learner}\label{sec: global collaborative learner}
Despite the considerable results achieved by intra-sequence modeling, the lack of global collaborative information obtained by inter-sequence exploitation will result in inferior results. Previous methods\cite{GCEGNN, DGNN, GCL4SR} have attempted to alleviate this issue by introducing a global item graph constructed with pre-defined rules. However, we argue that the rule-based graph may contain noisy edges or overlook underlying connections as in Fig. \ref{fig: motivation}, thus we aim to learn an adaptive global item graph and incorporate the refined global collaborative information into the sequential modeling in an SSL fashion. The detailed implementations are as follows.

\subsubsection{\textbf{Refined Graph Representation Learning}}\label{sec: refined graph representaion}
We construct a rule-based global item graph as the initial graph in Sec. \ref{sec: initial graph construction}. Then we intend to refine the graph with small perturbation, which is formalized in a direct approach fashion introduced in Sec. \ref{sec: GSL}:
\begin{equation}\label{eq: refined graph}
    \hat{\mathcal{A}} = \mathcal{A} + \alpha \mathcal{A}',
\end{equation}
where $\mathcal{A} \in \mathbb{R}^{|\mathcal{V}| \times |\mathcal{V}|}$ is the fixed original graph, $\mathcal{A}' \in \mathbb{R}^{|\mathcal{V}| \times |\mathcal{V}|}$ is a fully learnable perturbation graph, $\alpha$ is the strength of perturbation, and $\hat{\mathcal{A}} \in \mathbb{R}^{|\mathcal{V}| \times |\mathcal{V}|}$ is the refined graph, which will be used to generate graph representations. Specifically, we can leverage LightGCN\cite{LightGCN} in Sec. \ref{sec: LightGCN} to generate graph representations. We denote node embeddings (item embeddings) as $\mathbf{V} \in \mathbb{R}^{|\mathcal{V}| \times d}$, then we can obtain the original graph representation $\mathbf{E}^{(L)}$ by substituting $\mathbf{E}^{(0)}$ with $\mathbf{V}$, and the refined graph representation $\hat{\mathbf{E}}^{(L)}$ by further substituting $\mathcal{A}$ with $\hat{\mathcal{A}}$ in Eq. \ref{eq: refined graph}.

\subsubsection{\textbf{SVD-based Acceleration}}\label{sec: svd acceleration}
However, the computational complexity of Eq. \ref{eq: refined graph} is quadratic with respect to the number of items, which is unacceptable for large-scale recommendation systems. Therefore, we propose an SVD-based acceleration method to accelerate the learning process of the perturbed graph $\mathcal{A}'$. Specifically, we observe that smaller singular values of real-world item graphs account for the majority of all singular values, so we make the same assumption for the perturbed graph, which enables us to accelerate the modeling of $\mathcal{A}'$ as we only need to model a small portion of dominating singular values. In this way, instead of constructing the whole perturbed graph, we alternate to learn the SVD decomposition of it, which can be formalized as follows,
\begin{equation}\label{eq: SVD refined graph}
    \mathcal{A}' = (\mathcal{A} \mathbf{W}_{US})(\mathcal{A} \mathbf{W}_{V})^T,
\end{equation}
where $\mathbf{W}_{US} \in \mathbb{R}^{|\mathcal{V}| \times d'}$ and $\mathbf{W}_{V} \in \mathbb{R}^{|\mathcal{V}| \times d'}$ are both learnable parameters, $d'$ is a hyper-parameter used to control the rank of the decomposed perturbed graph. With the decomposition form of $\mathcal{A}'$, we can now re-write Eq. \ref{eq: LightGCN message passing} as,
\begin{equation}\label{eq: SVD-based message passing}
    \mathbf{E}^{(l)} = \mathcal{A} \mathbf{E}^{(l - 1)} + (\mathcal{A} \mathbf{W}_{US})(\mathcal{A} \mathbf{W}_{V})^T \mathbf{E}^{(l - 1)},
\end{equation}
where we can first compute $(\mathcal{A} \mathbf{W}_{V})^T \mathbf{E}^{(l - 1)}$, whose complexity is $O(|\mathcal{V}|dd')$, and then left multiply $\mathcal{A} \mathbf{W}_{US}$ onto the result, whose complexity is also $O(\mathcal{V}dd')$. If we choose a small $d' \ll |\mathcal{V}|$, we can reduce the time and space complexity from quadratic to linear.


\subsubsection{\textbf{Global Collaborative Information Encoding}}
Previous methods like GC-SAN\cite{GCSAN} directly feed the graph-enhanced item embedding $\mathbf{E}^{(L)}$ into the sequential encoder defined in \ref{sec: sequential modeling}, but we empirically find it ineffective. Thus, we instead leverage an SSL paradigm to incorporate the refined global collaborative information into item representations efficiently, which is accomplished through mutual information maximization \cite{DeepInfoMax, S3Rec}. Formally, the mutual information between two random variables $X$ and $Y$ is defined as:
\begin{equation}\label{eq: mutual information}
    I(X;Y) \equiv H(X) - H(X|Y) \equiv H(Y) - H(Y|X),
\end{equation}
where $H(X)$ and $H(Y)$ are the individual information entropy of $X$ and $Y$ respectively while $H(X|Y)$ and $H(Y|X)$ are conditional information entropy. Although it is intractable to directly maximize the mutual information, we can estimate it effectively by InfoNCE\cite{CPC}:
\begin{equation}\label{eq: InfoNCE}
    \mathcal{L}_{InfoNCE} = - \sum_{i=1}^K \log{\frac{e^{f(x_i, y_i)}}{\sum_{j=1}^K e^{f(x_i, y_j)}}},
\end{equation}
where $K$ is the number of samples and $f$ is a critic function. In this work, graph representation generated based on the original or refined graph should possess similar information, so we intend to maximize the mutual information between the original graph and the refined graph. Specifically, we compute Eq. \ref{eq: InfoNCE} within a batch, and instantiate $f$ as the cosine similarity between the original and refined graph representation, which can be formalized as follows:
\begin{equation}\label{eq: gce loss}
    \mathcal{L}_{gce} = - \sum_{i=1}^{|B|} \log{\frac
        {e^{cos(\mathbf{e}_i^{(L)}, \hat{\mathbf{e}}_i^{(L)}) / \tau}}
        {\sum_{j=1}^{|B|} e^{cos(\mathbf{e}_i^{(L)}, \hat{\mathbf{e}}_j^{(L)}) / \tau}}
    },
\end{equation}
where $\mathbf{e}_i^{(L)} \in \mathbf{E}^{(L)}$ and $\hat{\mathbf{e}}_i^{(L)} \in \hat{\mathbf{E}}^{(L)}$ are obtained in Sec. \ref{sec: refined graph representaion}, $\tau$ is a temprature hyper-parameter, which we fix to $0.2$ empirically.

\subsection{Sequential Modeling}\label{sec: sequential modeling}
With the global-information-enhanced item embeddings, now we can incorporate the adaptive global collaborative information into the sequential modeling process. Specifically, we follow previous works\cite{SASRec, CL4SRec, ICLRec} to adopt a Transformer\cite{transformer}-based encoder for its powerful capability. A Transformer-based encoder takes sequences of item embedding as input and consists of several multi-head self-attention (MHSA) layers and feed-forward (FFN) layers.

Considering a sequence $s_u = [v_1, v_2, \dots, v_t, \dots, v_{|s_u|}]$ defined in Sec. \ref{sec: next item prediction}, we first project them into a sequence of item embedding $[\mathbf{v}_1, \mathbf{v}_2, \dots, \mathbf{v}_t, \dots, \mathbf{v}_{|s_u|}]$, where $\mathbf{v}_t$ is the embedding vector of item $v_t$. Then we add a learnable positional encoding $\mathbf{P} \in \mathbb{R}^{N \times d}$ onto the item embeddings to obtain the input of Transformer layers $\mathbf{H}^{(0)}$. Moreover, given the output $\mathbf{H}^{(l - 1)} \in \mathbb{R}^{N \times d}$ at the $(l - 1$)-th layer, output at the $l$-th layer $\mathbf{H}^{(l)}$ can be obtained by:
\begin{equation}
    \mathbf{H}^{(l)} = FFN(MHSA(\mathbf{H}^{(l - 1)})),
\end{equation}
where a single head of MHSA is defined as follows,
\begin{equation}\label{eq: attention}
    Attention(\mathbf{Q}, \mathbf{K}, \mathbf{V}) = \left( softmax \left(\frac{\mathbf{Q}\mathbf{K}^T}{\sqrt{d}} \right) \right)\mathbf{V},
\end{equation}
where $\mathbf{Q}$, $\mathbf{K}$, and $\mathbf{V}$ are the query, key, and value matrix of the Transformer layers. Assuming the output of the last layer $L$ is $\mathbf{H}^{(L)} = [\mathbf{h}_1^{(L)}, \mathbf{h}_2^{(L)}, \dots, \mathbf{h}_{|s_u|}^{(L)}]$, we treat the last hidden vector $\mathbf{h}_{|s_u|}^{(L)}$ as users' preference representations.

With the generated user representations and the item embeddings, the sequential model can be optimized by the next item prediction loss as follows,
\begin{equation}\label{eq: next item loss}
    \mathcal{L}_{rec} = \sum_{u = 1}^{|\mathcal{U}|} \sum_{t=2}^{|s_u|} (-\log(\sigma(\mathbf{h}_{t-1}^u \cdot \mathbf{v}_t^u)) - \sum_{v_j \notin s_u} \log(1 - \sigma(\mathbf{h}_{t-1}^u \cdot \mathbf{v}_j))),
\end{equation}
where $\mathbf{h}_{t-1}^u$ is the preference representation of user $u$ at time step $t - 1$, $\mathbf{v}_t^u$ is the embedding of the target item in $s_u$, and we randomly sample one negative item $v_j$ at each time step $t$ for each sequence. The purpose of Eq. \ref{eq: next item loss} is to maximize the probability of recommending the expected next item $v_t^u$ to the user $u$ given the user's preference representation $\mathbf{h}_{t-1}^u$, which is obtained from the previous sequences $[v_1, v_2, \dots, v_{t-1}]$.

\subsection{Personalized Graph Extractor}\label{sec: graph adaption}
The graph introduced in Sec. \ref{sec: global collaborative learner} focuses on encoding global collaborative information, reflecting general connections among items. But in the end, we aim to model the item correlation for each specific user. So we propose to extract personalized information from the global graph. Previous work\cite{GCL4SR} fuses the personalized information with the sequential information in the representation level. However, we empirically find it less effective and we suspect it is caused by the incompatibility between graph and sequential representations. Therefore, we intend to maintain two required properties of the extracted information: \textbf{compatibility} and \textbf{personalization}.

Considering compatibility, as the self-attention mechanism can also be regarded as learning edge weights between item pairs within a sub-graph, we propose to inject the personalized global collaborative information into the sequential encoder in the form of relative positional encoding, with which the attention computation in Eq. \ref{eq: attention} can be formalized as follows,
\begin{equation}\label{eq: attention with relative pe}
    Attention(\mathbf{Q}_u, \mathbf{K}_u, \mathbf{V}_u) = \left( softmax \left(\frac{\mathbf{Q}_u\mathbf{K}^T_u}{\sqrt{d}} + \mathbf{P}'_u \right) \right)\mathbf{V}_u,
\end{equation}
where $u \in \mathcal{U}$ is some user, and $\mathbf{P}'_u \in \mathbb{R}^{N \times N}$ is the relative positional encoding matrix generated for this user.

Then we need to specify the relative positional encoding for the consideration of personalization. To meet the requirement, we introduce a personalized embedding vector $\mathbf{s}_u \in \mathbb{R}^{1 \times d}$ for each user. Specifically, given the user $u$'s interaction sequence $s_u = [v_1, v_2, \dots, v_t, \dots, v_N]$, we can extract a sub-graph $\tilde{\mathcal{A}} \in \mathbb{R}^{N \times N}$ from the global item graph according to the interacted items in $s_u$, and then we can obtain the personalized relative positional encoding for the user $u$ as follows,
\begin{equation}\label{eq: personalized relative pe}
    \mathbf{P}'_u = MLP(\mathbf{s}_u) \tilde{\mathcal{A}},
\end{equation}
where $MLP(\cdot)$ is used to project the user embedding vector into a scalar, measuring the importance $u$ places on global information.

Through combining Eq. \ref{eq: attention with relative pe} and Eq. \ref{eq: personalized relative pe}, we can extract personalized information from the global graph for each user and integrate them with the sequential encoder in a highly compatible way.
\subsection{Model Optimization}\label{sec: multi-task learning}
\subsubsection{\textbf{Learning Objective}}
We leverage a multi-task learning paradigm to simultaneously optimize the entire framework:
\begin{equation}\label{eq: total loss}
    \mathcal{L}_{total} = \mathcal{L}_{rec} + \lambda_1 \mathcal{L}_{gce} + \lambda_2 \mathcal{L}_{seq},
\end{equation}
where $\mathcal{L}_{rec}$ is the next item prediction loss in Eq. \ref{eq: next item loss}, $\mathcal{L}_{gce}$ is the global collaborative encoding loss in Eq. \ref{eq: gce loss}, $\mathcal{L}_{seq}$ is the sequence-level contrastive learning task introduced in CL4SRec\cite{CL4SRec}, which augments sequences by cropping, masking or reordering, and $\lambda_1$ and $\lambda_2$ are hyper-parameters that are used to control the strength of the two SSL objectives respectively.

\subsubsection{\textbf{Computational Complexity}}
The computational complexity mainly comes from the adaptive global collaborative learner, so we only discuss its complexity for brevity. The complexity of the message-passing operation of LightGCN in Eq. \ref{eq: LightGCN message passing} is $O(|\mathcal{E}|d)$, where $|\mathcal{E}|$ is the number of edges in the graph $\mathcal{A}$. With the SVD-acceleration strategy in Sec. \ref{sec: svd acceleration}, we can reduce the computational complexity of learning a perturbation graph from $O(|\mathcal{V}|^2)$ to $O(|\mathcal{V}|dd')$. And the complexity of the InfoNCE loss in Eq. \ref{eq: gce loss} is $O(|B|^2d)$. Therefore, the entire complexity of the adaptive global collaborative learner is $O(|\mathcal{E}|d + |\mathcal{V}|dd' + |B|^2d)$. As we can choose $d' \ll |\mathcal{V}|$ and a small batch size $B$, the final computational complexity is linear with respect to the number of interaction records.

\section{EXPERIMENTAL EVALUATION}
\subsection{Experimental Settings}
\subsubsection{\textbf{Datasets}}

\begin{table}
\centering
\caption{Statistics of the datasets.}
\label{tab: dataset}
\begin{tabular}{l|llll} 
\hline
Dataset         & Beauty  & Sports  & Toys    & Yelp     \\ 
\hline
$|\mathcal{U}|$ & 22,363  & 35,598  & 19,412  & 30,431   \\
$|\mathcal{V}|$ & 12,101  & 18,357  & 11,924  & 20,033   \\
\# Interactions & 0.2m    & 0.3m    & 0.17m   & 0.3m     \\
Avg. length     & 8.9     & 8.3     & 8.6     & 8.3      \\
Sparsity        & 99.95\% & 99.95\% & 99.93\% & 99.95\%  \\
\hline
\end{tabular}
\vspace{-0.3cm}
\end{table}

To validate the effectiveness of our proposed method, we conduct experiments on four publicly available and widely adopted datasets:
\begin{itemize}[leftmargin=*]
\item \textbf{Beauty, Sports, Toys\footnote{\url{http://snap.stanford.edu/data/amazon/productGraph/categoryFiles/}}}: Introduced in \cite{amazon_dataset}, the Amazon-review data is collected from one of the biggest e-commerce website Amazon and widely used in evaluating recommendation systems. We select three categories "Beauty", "Sports and Outdoors" and "Toys and Games" from the collection.
\item \textbf{Yelp\footnote{\url{https://www.yelp.com/dataset}}}: This is a business review dataset that treats restaurants, bars, and other local businesses as items.
\end{itemize}

As for preprocessing the datasets, we follow \cite{CL4SRec, ICLRec} to ensure each user and item has at least 5 associated interactions. Afterward, we group the interaction records by users and sort each interaction group by the timestamp to get each user's interaction sequence. Statistics of the preprocessed datasets are summarized in Table \ref{tab: dataset}.

\subsubsection{\textbf{Evaluation Protocols}}
To evaluate the top-K recommendation performance, we consider the commonly used leave-one-out strategy as \cite{SASRec, CL4SRec, ICLRec}. Specifically, we utilize the most recent interaction in each user's sequence as the test data, the second most recent interaction as the validation data, and all the remaining interactions as training data. As for the evaluation metrics, we employ HR and NDCG, which are used in previous works\cite{SASRec, CL4SRec, ICLRec}. Besides, following the suggestion of Krichene and Rendle\cite{sampled_metrics}, we adopt the whole item set as the candidate item set during evaluation to avoid the sampling bias of the candidate selection.

\subsubsection{\textbf{Compared Baselines}}
We compare the performance of our method with several kinds of state-of-the-art baselines, whose detailed information can be found in Sec. \ref{sec: sequential modeling}.

\begin{itemize}[leftmargin=*]
\item \textbf{Non-sequential methods}: BPR-MF\cite{BPR} is a simple yet effective matrix factorization method with a pair-wise Bayesian Personalized Ranking (BPR) loss function. 
\item \textbf{General sequential methods}: We include a CNN-based method Caser\cite{Caser}, a RNN-based method GRU4Rec\cite{GRU4Rec} and a Transformer-based method SASRec\cite{SASRec}.
\item \textbf{Graph-based sequential methods}: We choose two local-graph-based baselines SR-GNN\cite{SRGNN} and GC-SAN\cite{GCSAN} as our baselines. As for global-graph-based approaches, we choose GCE-GNN\cite{GCEGNN} and a GSL-based method DGNN\cite{DGNN}.
\item \textbf{SSL-based sequential methods}: As the main competitors of APGL4SR, we select SSL-based baselines from multiple perspectives. CL4SRec\cite{CL4SRec} is a sequence-level SSL baseline, DuoRec\cite{DuoRec} performs contrastive learning in the model level, ICLRec\cite{ICLRec} proposes an intent-level SSL paradigm, and GCL4SR\cite{GCL4SR} introduces graph contrastive learning for graph-level SSL.
\end{itemize}

\subsubsection{\textbf{Implementation Details}}
We implement SASRec, CL4SRec, ICLRec, and APGL4SR based on the framework provided by ICLRec\footnote{\url{https://github.com/salesforce/ICLRec}}, other baselines are implemented with RecStudio\cite{recstudio} and RecBole-GNN\cite{recbole2}. The maximum number of training epochs is set to 1000 and the training procedure will be early stopped when NDCG@20 on the validation data doesn't increase for 40 epochs. For all methods, we adopt the following settings to perform a fair comparison: the training batch size $B$ is fixed to 256, Adam\cite{Adam} with a learning rate of 1e-3 is used as the optimizer, the embedding size $d$ is set to 64 and the maximum sequence length $N$ is set to 50 for all datasets. For each baseline, all other hyper-parameters are searched as the original papers suggest. Notably, we utilize BCE loss in Eq. \ref{eq: next item loss} as the next item prediction loss for all the methods rather than CE loss adopted by previous methods\cite{DuoRec, GCL4SR} for fair comparisons.

Considering particular hyper-parameters of APGL4SR, we set the number of LightGCN layers $L$ and perturbation strength $\alpha$ in Eq. \ref{eq: LightGCN final representation} as 2 and 0.05 respectively, and we set $d'$ in Eq. \ref{eq: SVD refined graph} to 256 for efficiency. As for the sequential encoder, the number of attention heads is set to 2 and the number of layers is searched among $[1, 2, 3]$. As for the strength of SSL tasks, we search the strength of global collaborative encoding loss $\lambda_1$ among $[0.05, 0.1, 0.2, 0.4]$ and fix the strength of sequence-level SSL loss $\lambda_2$ to 0.1.

\subsection{Overall Performance}

\begin{table*}[]
\centering
\caption{The overall performance of all baselines. The best result is bolded while the second-best result is underlined in each row. We run the proposed method and the most competitive baselines with ten different random seeds and analyze statistical significance, where * represents p-value < 0.01, ** represents p-value < 0.001.}
\label{tab: overall performance}
\vspace{-0.2cm}
\setlength\tabcolsep{1.5pt}
\resizebox{\textwidth}{!}{
\begin{tabular}{c|l|c|ccc|cccc|cccc|c|c}
\hline
Dataset                 & Metrics & BPR    & Caser  & GRU4Rec & SASRec & SR-GNN & GC-SAN & GCE-GNN & DGNN   & CL4SRec & GCL4SR & DuoRec & ICLRec        & APGL4SR                & Improv \\ \hline
\multirow{4}{*}{Beauty} & HR@5    & 0.0191 & 0.0228 & 0.0190  & 0.0356 & 0.0231 & 0.0309 & 0.0260  & 0.0371 & 0.0453  & 0.0415 & 0.0449 & \underline{0.0496}$\pm$0.0010 & \textbf{0.0543}$\pm$0.0010 & 9.47\%** \\
                        & HR@20   & 0.0527 & 0.0601 & 0.0568  & 0.0855 & 0.0609 & 0.0798 & 0.0653  & 0.0876 & 0.1045  & 0.1012 & 0.1039 & \underline{0.1059}$\pm$0.0013 & \textbf{0.1093}$\pm$0.0011 & 3.19\%** \\
                        & NDCG@5  & 0.0119 & 0.0135 & 0.0119  & 0.0227 & 0.0133 & 0.0163 & 0.0157  & 0.0240 & 0.0294  & 0.0263 & 0.0284 & \underline{0.0323}$\pm$0.0006 & \textbf{0.0372}$\pm$0.0007 & 12.8\%** \\
                        & NDCG@20 & 0.0218 & 0.0272 & 0.0224  & 0.0373 & 0.0264 & 0.0302 & 0.0286  & 0.0392 & 0.0461  & 0.0421 & 0.0453 & \underline{0.0480}$\pm$0.0006 & \textbf{0.0527}$\pm$0.0006 & 8.07\%** \\ \hline
\multirow{4}{*}{Sports} & HR@5    & 0.0129 & 0.0154 & 0.0110  & 0.0183 & 0.0152 & 0.0161 & 0.0154  & 0.0197 & 0.0261  & 0.0233 & 0.0265 & \underline{0.0272}$\pm$0.0005 & \textbf{0.0299}$\pm$0.0007 & 9.56\%** \\
                        & HR@20   & 0.0344 & 0.0399 & 0.0289  & 0.0450 & 0.0405 & 0.0437 & 0.0425  & 0.0470 & 0.0611  & 0.0571 & 0.0615 & \underline{0.0637}$\pm$0.0007 & \textbf{0.0664}$\pm$0.0009 & 4.30\%** \\
                        & HR@5    & 0.0091 & 0.0114 & 0.0065  & 0.0135 & 0.0075 & 0.0084 & 0.0082  & 0.0139 & 0.0166  & 0.0145 & 0.0169 & \underline{0.0179}$\pm$0.0002 & \textbf{0.0201}$\pm$0.0006 & 12.0\%** \\
                        & HR@20   & 0.0136 & 0.0178 & 0.0115  & 0.0186 & 0.0153 & 0.0162 & 0.0159  & 0.0203 & 0.0263  & 0.0232 & 0.0267 & \underline{0.0281}$\pm$0.0002 & \textbf{0.0304}$\pm$0.0007 & 8.11\%** \\ \hline
\multirow{4}{*}{Toys}   & HR@5    & 0.0181 & 0.0142 & 0.0178  & 0.0431 & 0.0282 & 0.0417 & 0.0312  & 0.0445 & 0.0535  & 0.0501 & 0.0542 & \underline{0.0577}$\pm$0.0005 & \textbf{0.0627}$\pm$0.0009 & 8.64\%** \\
                        & HR@20   & 0.0495 & 0.0431 & 0.0467  & 0.0886 & 0.0645 & 0.0863 & 0.0721  & 0.0921 & 0.1098  & 0.1042 & 0.1121 & \underline{0.1136}$\pm$0.0010 & \textbf{0.1176}$\pm$0.0012 & 3.43\%** \\
                        & NDCG@5  & 0.0135 & 0.0094 & 0.0114  & 0.0283 & 0.0191 & 0.0253 & 0.0223  & 0.0291 & 0.0365  & 0.0326 & 0.0372 & \underline{0.0393}$\pm$0.0005 & \textbf{0.0433}$\pm$0.0005 & 10.1\%** \\
                        & NDCG@20 & 0.0225 & 0.0172 & 0.0194  & 0.0409 & 0.0292 & 0.0382 & 0.0343  & 0.0421 & 0.0528  & 0.0487 & 0.0537 & \underline{0.0551}$\pm$0.0003 & \textbf{0.0588}$\pm$0.0006 & 6.72\%** \\ \hline
\multirow{4}{*}{Yelp}   & HR@5    & 0.0112 & 0.0137 & 0.0129  & 0.0160 & 0.0117 & 0.0150 & 0.0121  & 0.0166 & 0.0227  & 0.0204 & 0.0215 & \underline{0.0239}$\pm$0.0005 & \textbf{0.0248}$\pm$0.0005 & 3.82\%* \\
                        & HR@20   & 0.0371 & 0.0401 & 0.0369  & 0.0437 & 0.0375 & 0.0417 & 0.0382  & 0.0452 & 0.0629  & 0.0587 & 0.0621 & \underline{0.0650}$\pm$0.0004 & \textbf{0.0670}$\pm$0.0003 & 3.07\%** \\
                        & NDCG@5  & 0.0084 & 0.0088 & 0.0078  & 0.0101 & 0.0087 & 0.0096 & 0.0091  & 0.0105 & 0.0143  & 0.0121 & 0.0137 & \underline{0.0150}$\pm$0.0003 & \textbf{0.0157}$\pm$0.0002 & 4.43\%* \\
                        & NDCG@20 & 0.0143 & 0.0152 & 0.0145  & 0.0177 & 0.0148 & 0.0171 & 0.0157  & 0.0180 & 0.0255  & 0.0214 & 0.0246 & \underline{0.0264}$\pm$0.0001 & \textbf{0.0274}$\pm$0.0002 & 3.66\%** \\ \hline
\end{tabular}}
\end{table*}

In this subsection, we compare the overall performance of all methods, which is presented in Table \ref{tab: overall performance}. We can draw the following conclusions from the results: (1) By comparing BPR-MF and other sequential methods, we can find BPR-MF basically yields worse results, indicating the significance of capturing dynamic user preferences. (2) By comparing graph-based SR methods, we can observe that GCE-GNN outperforms SR-GNN, which can be attributed to the introduced global collaborative information. Though GC-SAN can yield better results than GCE-GNN, this is mainly due to it adopting a Transformer-based encoder. So we can observe that DGNN can beat GC-SAN because it considers global information and stronger sequential encoders simultaneously. (3) By comparing SSL-based methods and non-SSL methods, we can observe that SSL-based consistently perform better than non-SSL baselines, which mainly comes from the effectiveness of SSL in alleviating the data-sparsity issue. (4) By comparing the proposed APGL4SR with other SSL-based methods, APGL4SR consistently outperforms other baselines on all datasets. On the one hand, our method incorporates global collaborative information into the sequential modeling process, yielding better user representations than CL4SRec, DuoRec, and ICLRec. On the other hand, the proposed two modules realize the adaptive construction and compatible utilization of the global collaborative graph respectively, which generates more refined global collaborative information than GCL4SR.


\subsection{Ablation Study}\label{sec: ablation study}

\begin{table}[]
\caption{Abalation study of APGL4SR on NDCG@20.}
\vspace{-0.2cm}
\label{tab: ablation study}
\begin{tabular}{l|cccc}
\hline
Model                       & Beauty               & Sports               & Toys                 & Yelp                 \\ \hline
(A) APGL4SR                                 & \textbf{0.0538}               & \textbf{0.0308}               & \textbf{0.0584}               & \textbf{0.0275}               \\
(B) w/o AGC                                 & 0.0470               & 0.0271               & 0.0535               & 0.0261               \\
(C) w/o PGE                                 & 0.0526               & 0.0297               & 0.0573               & 0.0267               \\ \hline
(D) $\mathcal{A}_{SVD}$                                     & 0.0089               & 0.0079               & 0.0525               & 0.0248               \\
(E) Fusion                                  & 0.0370               & 0.0195               & 0.0398               & 0.0178               \\ \hline
\end{tabular}
\vspace{-0.3cm}
\end{table}

In this subsection, we intend to demonstrate the effectiveness of each module of APGL4SR. The results are presented in Table \ref{tab: ablation study}, wherein (B) removes the adaptive global collaborative learner while (C) removes the personalized graph extractor. By comparing (A) and (B), we can observe that the performance can be greatly improved by incorporating global collaborative information into the initial item representations, demonstrating the importance of modeling global collaborative information. By comparing (A) and (C), we can find the personalized global collaborative information injected into the sequential encoder can help it to learn better user representation. 

Besides, we aim to prove the effectiveness of APGL4SR compared with existing graph construction and utilization strategies, so we further introduce some variants in Table \ref{tab: ablation study}. We notice that LightGCL\cite{LightGCL} proposes to directly contrast representations from the original graph and a reconstructed graph by SVD, so we add a variant (D) that replaces $\mathcal{A}'$ in Eq. \ref{eq: refined graph} with a fixed graph reconstructed from SVD decomposition. Moreover, to validate the importance of compatibility, we further introduce a variant (E) that directly fuses graph representation and sequential representation with an MLP instead of the graph extractor. By comparing (A) and (D), we can find the performance of (D) drops severely in several datasets. The reason might be that our refined graph can be jointly optimized, by which we can obtain a more adaptive graph to incorporate the global collaborative information. Besides, by comparing (A) and (E), we can observe a significant performance drop when we directly fuse the graph and sequential representations, demonstrating the necessity of compatibility of personalized graph extraction.

\subsection{Hyper-parameter Sensitivity}
We investigate the hyper-parameter sensitivity of our method in this section. Specifically, we focus on how the strength of global collaborative encoding loss $\lambda_1$ in Eq. \ref{eq: total loss} and the strength of graph perturbation in Eq. \ref{eq: refined graph} influence the recommendation performance. We only report the results on Beauty and Toys dataset for simplicity.
\subsubsection{\textbf{Strength of Global Collaborative Encoding $\lambda_1$}}

\begin{figure}[t!]
    \centering
    \begin{subfigure}[t]{0.23\textwidth}
           \centering
           \includegraphics[width=\textwidth]{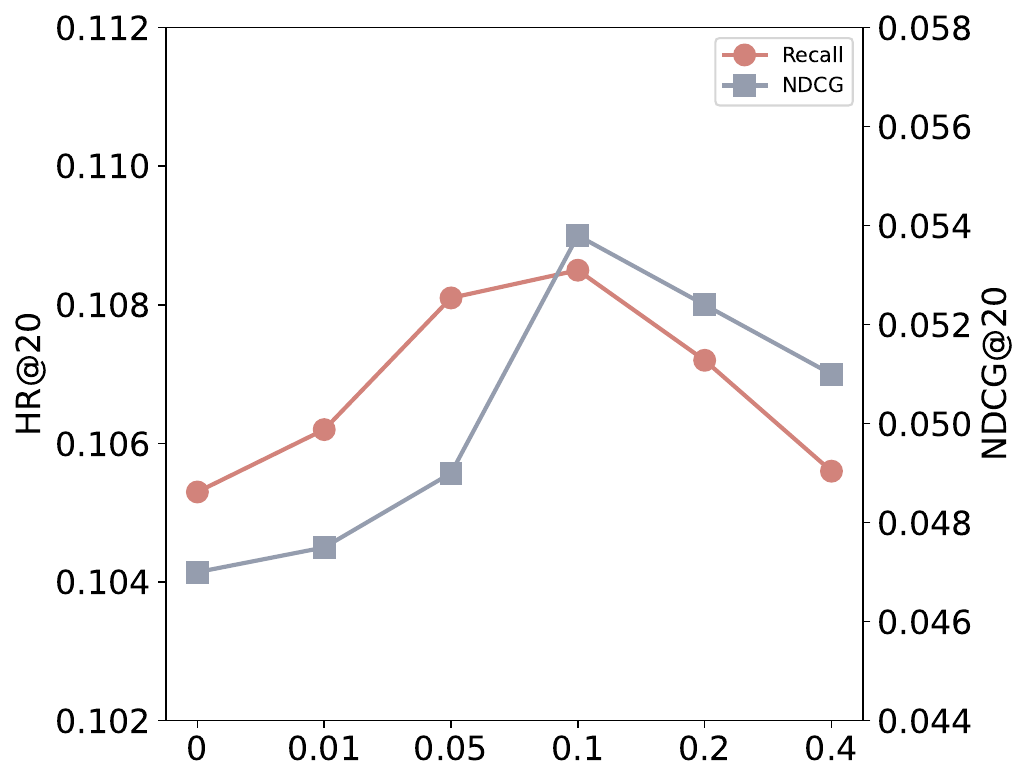}
            \caption{Beauty}
            \label{fig:lambda_sensitivity_Beauty}
    \end{subfigure}
    \begin{subfigure}[t]{0.23\textwidth}
            \centering
            \includegraphics[width=\textwidth]{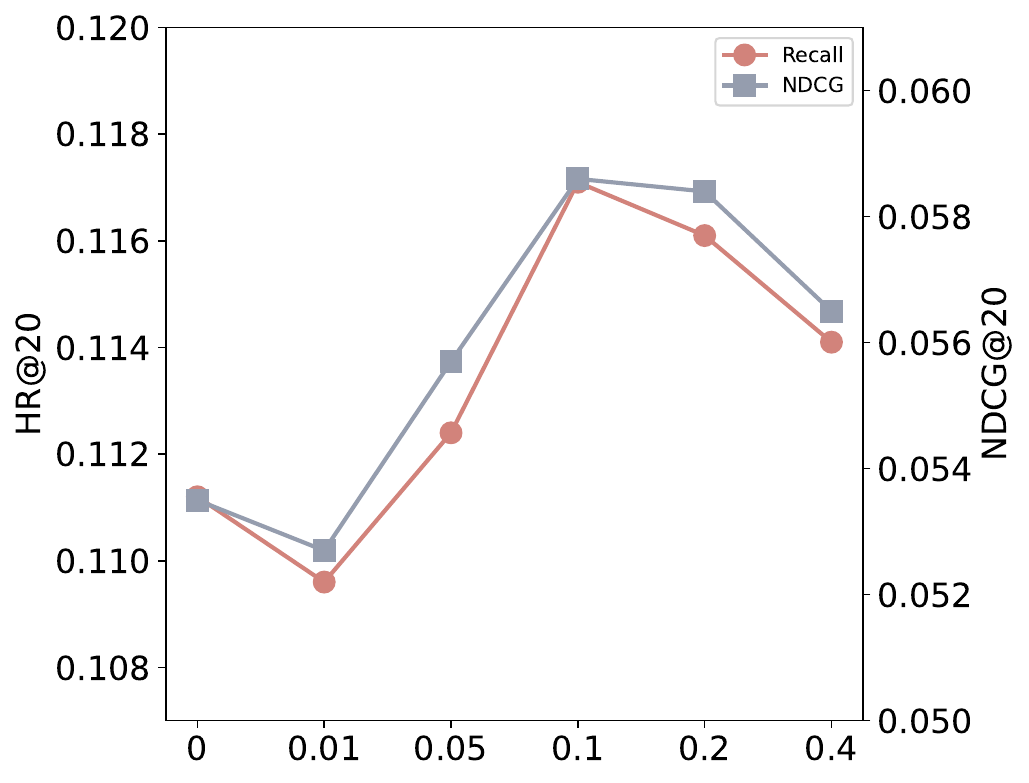}
            \caption{Toys}
            \label{fig:lambda_sensitivity_Toys}
    \end{subfigure}
    \vspace{-0.3cm}
    \caption{Recommendation Performance w.r.t $\lambda_1$}
    \label{fig: lambda analysis}    
    \vspace{-0.3cm}
\end{figure}

We set $\lambda_1$ in the range of $[0, 0.01, 0.05, 0.1, 0.2, 0.4]$ and the results are depicted in Fig. \ref{fig: lambda analysis}. In the figure, we can observe that the performance exhibits an increasing trend as $\lambda_1$ increases because the global collaborative information incorporated into item embeddings can provide useful information for subsequent sequential modeling. However, the performance begins to decrease when $\lambda_1$ exceeds a threshold. The reason is that the large $\lambda_1$ will force item embeddings to encode excessive global collaborative information, instead preventing sequential signal modeling in subsequent modules.

\subsubsection{\textbf{Strength of Graph Perturbation $\alpha$}}

\begin{figure}[t!]
    \centering
    \begin{subfigure}[t]{0.23\textwidth}
           \centering
           \includegraphics[width=\textwidth]{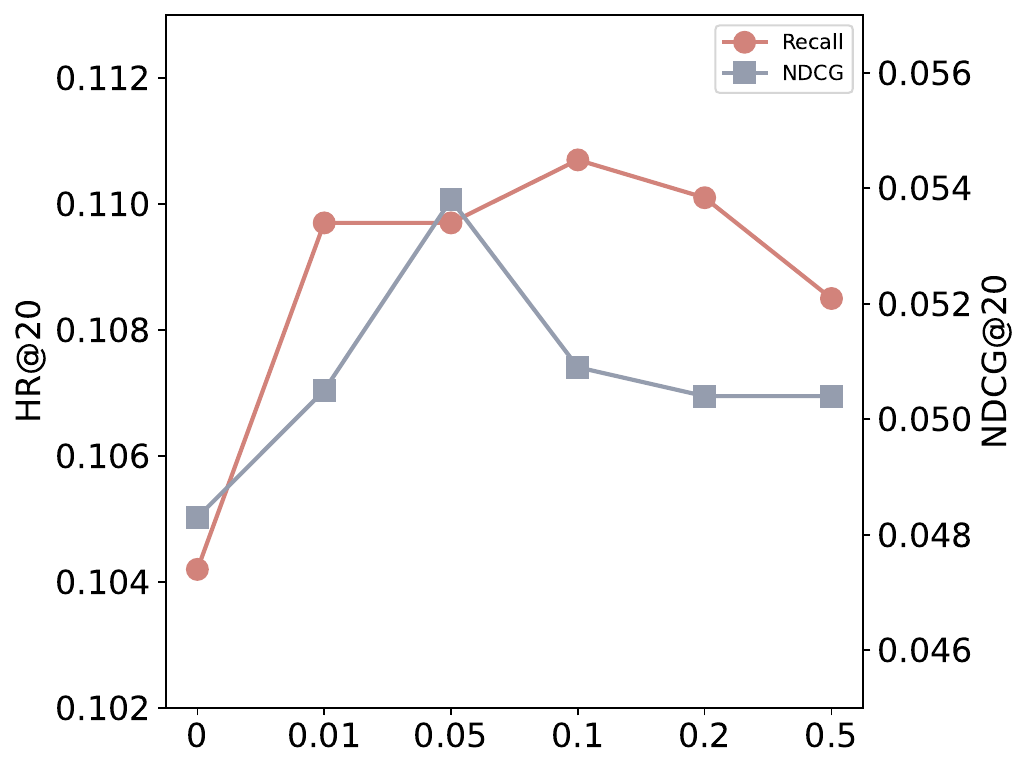}
            \caption{Beauty}
            \label{fig:alpha_sensitivity_Beauty}
    \end{subfigure}
    \begin{subfigure}[t]{0.23\textwidth}
            \centering
            \includegraphics[width=\textwidth]{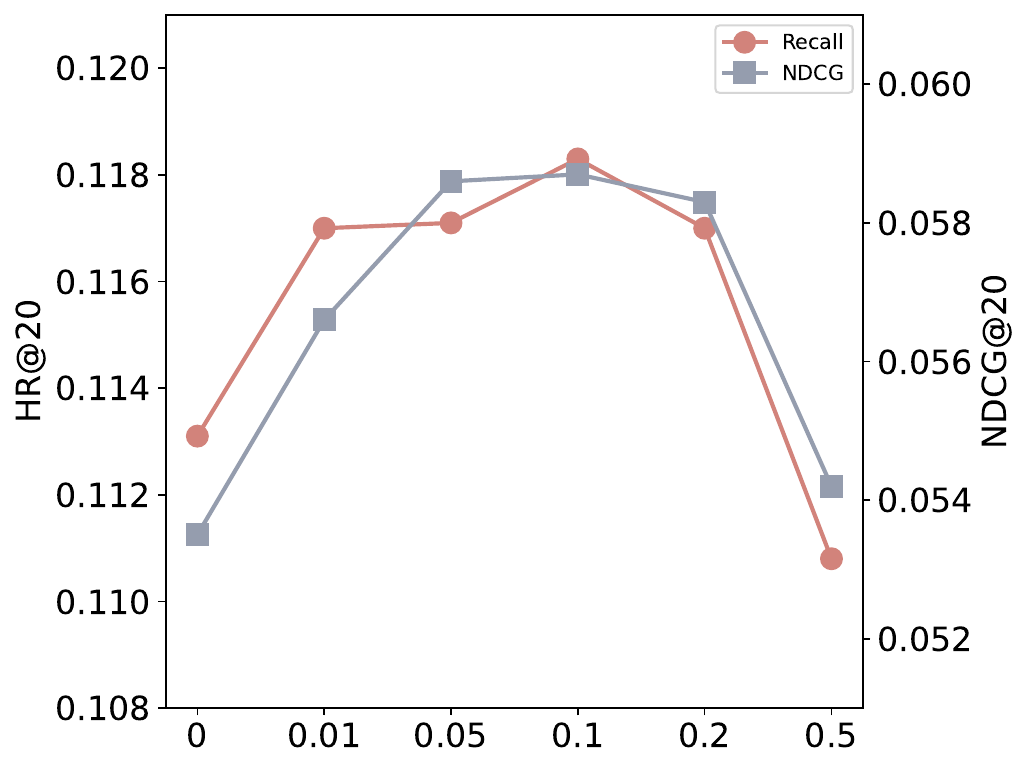}
            \caption{Toys}
            \label{fig:alpha_sensitivity_Toys}
    \end{subfigure}

    \caption{Recommendation Performance w.r.t $\alpha$}
    \label{fig: alpha analysis}
    \vspace{-0.6cm}
\end{figure}

We set $\alpha$ in the range of $[0, 0.01, 0.05, 0.1, 0.2, 0.5]$, and the results are illustrated in Fig. \ref{fig: alpha analysis}. We can observe the performance increases at first and then decreases, which means that we should control the strength of the graph perturbation. When $\alpha$ becomes too small, the refined graph is not flexible enough to provide informative training signals. However, too large $\alpha$ will make the refined graph fluctuate greatly, impairing the final performance.

\subsection{Further Experiments}
In this subsection, we will conduct some in-depth experiments to further verify and explain the effectiveness of the proposed method. First, we intend to demonstrate the versatility of the proposed framework. Then we verify the effectiveness of the refined graph in Eq. \ref{eq: refined graph}. Finally, we illustrate the superiority of our method in alleviating the dimensional collapse issue of item embeddings.

\subsubsection{\textbf{Analysis on Versatility of APGL4SR}}

\begin{table}[]
\caption{Analysis of versatility of APGL4SR on NDCG@20.}
\label{tab: versatility study}
\vspace{-0.2cm}
\begin{tabular}{l|cccc}
\hline
Model                       & Beauty          & Sports          & Toys            & Yelp            \\ \hline
(A) GRU4Rec                                 & 0.0224          & 0.0115          & 0.0194          & 0.0145          \\
(B) GRU4Rec+AGL                             & 0.0290          & 0.0152          & 0.0254          & 0.0180          \\ \hline
(C) SASRec                                  & 0.0373          & 0.0186          & 0.0409          & 0.0177          \\
(D) SASRec+APGL                             & 0.0452          & 0.0244          & 0.0513          & 0.0195          \\ \hline
(E) ICLRec                                  & 0.0480          & 0.0281          & 0.0551          & 0.0264          \\
(F) ICLRec+APGL                             & 0.0521          & 0.0290          & 0.0581          & 0.0273          \\ \hline
(G) APGL4SR                                 & \textbf{0.0538} & \textbf{0.0308} & \textbf{0.0584} & \textbf{0.0275} \\ \hline
\end{tabular}
\vspace{-0.4cm}
\end{table}

As a generic framework, the proposed modules can be easily applied to other sequential models. So we apply the two modules to other sequential methods to demonstrate the effectiveness of the proposed method. The corresponding results are presented in Table \ref{tab: versatility study}, where (B) is GRU4Rec with the global collaborative learner, while (D) and (F) are models equipped with both of the proposed modules.

In Table \ref{tab: versatility study}, from (A) to (F) we can find that the proposed modules can greatly improve the performance of the backbone model, which further confirms the effectiveness of APGL4SR. Besides, we can observe (F) and (G) produce comparable results, which is probably because the knowledge learned by ICLRec is a subset of knowledge obtained by APGL4SR. Moreover, (E) and (G) are both built upon CL4SRec, but our method achieves better performance, which is because APGL4SR can capture more global collaborative information complementing the sequence-level SSL task in CL4SRec.

\subsubsection{\textbf{Analysis on Effectiveness of the Refined Graph}}

\begin{table}[]
\centering
\caption{Analysis of the effectiveness of the adaptive graph.}
\vspace{-0.2cm}
\label{tab: graph analysis}
\begin{tabular}{l|cccc}
\hline 
 Model                      & Beauty          & Sports          & Toys            & Yelp            \\ \hline
(A) APGL4SR                                 & \textbf{0.0538} & \textbf{0.0308} & 0.0584          & \textbf{0.0275} \\
(B) FPG                                     & 0.0532          & 0.0291          & \textbf{0.0599} & 0.0271          \\
(C) NP                                      & 0.0483          & 0.0272          & 0.0535          & 0.263           \\ \hline
\end{tabular}
\vspace{-0.5cm}
\end{table}

To verify the effectiveness of the refined graph, we first train an APGL4SR model from scratch and extract the learned refined graph from the saved model. Then we train a new APGL4SR model by replacing the learnable refined graph in Eq. \ref{eq: refined graph} with the extracted refined graph. The corresponding results are presented in Table \ref{tab: graph analysis}, where (B) is the one with a fixed perturbation graph, and (C) means setting $\alpha$ to 0 so the graph is not perturbed. The table shows that (A) and (B) both outperform (C), verifying the importance of learning a more adaptive global collaborative graph. Besides, even with the graph fixed, we can find that (B) can still yield comparable or even better results compared with (A), which demonstrates the effectiveness of the learned refined graph, i.e., we can learn a more adaptive graph along with the training of the adaptive global collaborative learner.

\subsubsection{\textbf{Analysis on Avoiding Dimensional Collapse}}

\begin{figure}[t!]
    \centering
    \begin{subfigure}[t]{0.16\textwidth}
           \centering
           \includegraphics[width=\textwidth]{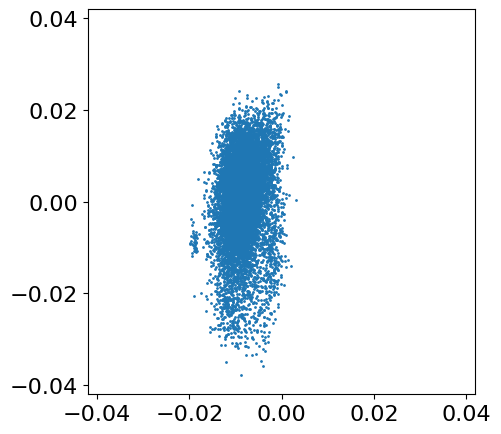}
            \caption{SASRec}
    \end{subfigure}
    \begin{subfigure}[t]{0.16\textwidth}
            \centering
            \includegraphics[width=\textwidth]{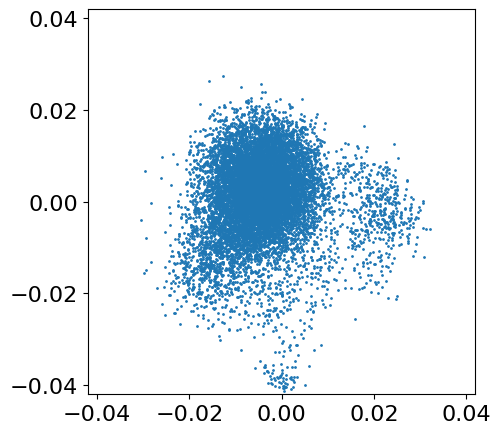}
            \caption{CL4SRec}
    \end{subfigure}
    \begin{subfigure}[t]{0.16\textwidth}
            \centering
            \includegraphics[width=\textwidth]{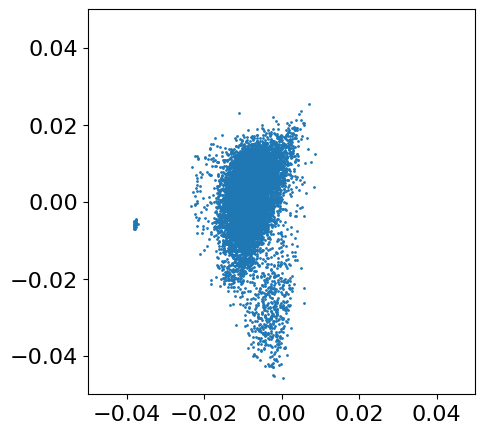}
            \caption{ICLRec}
    \end{subfigure}
    \begin{subfigure}[t]{0.16\textwidth}
            \centering
            \includegraphics[width=\textwidth]{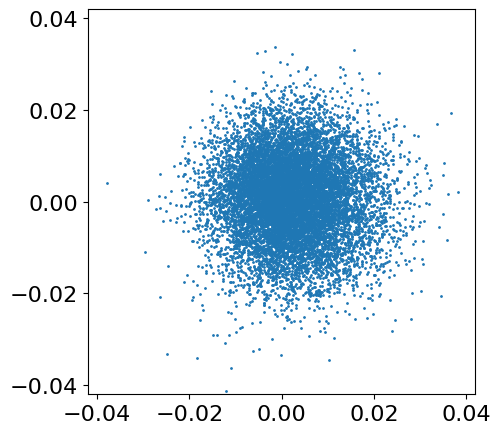}
            \caption{APGL4SR}
    \end{subfigure}

    \caption{Distribution visualizations of item embeddings.}
    \label{fig: distribution}
\vspace{-0.7cm}
\end{figure}

InfoNCE loss is known to promote the alignment and uniformity of representations on the hyper-sphere\cite{uniformity_and_alignment}. The uniformity of representations prefers a feature distribution preserving maximal information, which can greatly alleviate the dimensional collapse issue, i.e., the embedding vectors only span a lower-dimensional subspace instead of the entire available embedding space\cite{dimensional_collapse}. In this subsection, we intend to validate the superiority of our method in avoiding the dimensional collapse of the item embeddings, so we project the item embeddings learned on Beauty to 2D by SVD as in \cite{DuoRec}.

The visualizations of item embeddings are plotted in Fig. \ref{fig: distribution}. From Fig. \ref{fig: distribution}, we can observe item embeddings of SASRec coalesce in the latent space, resulting in less discriminative representations. By introducing some SSL tasks, item embeddings of CL4SRec achieve better uniformity. Surprisingly, we find ICLRec exhibits worse uniformity than CL4SRec though it yields better results. We suspect the reason may be that there exists meaningless uniformity in representations of CL4SRec, i.e., some relevant embeddings are unexpectedly pushed away. In contrast, our method can achieve more meaningful uniformity with the incorporation of adaptive global collaborative information, alleviating the dimensional collapse issue.

\section{CONCLUSIONS}
In this paper, we proposed a generic framework named APGL4SR to incorporate adaptive and personalized global collaborative information into sequential recommendation systems. We designed a novel learner to incorporate adaptive global collaborative information into item representations effectively and efficiently. Then we proposed a highly compatible personalized graph extractor to extract personalized information from the learned global graph for each user. Furthermore, we utilize a multi-task learning framework to train the entire model in an end-to-end fashion. As a generic framework, APGL4SR can not only outperform other baselines with significant margins on four public datasets, but also exhibit promising versatility, the ability to learn a meaningful global collaborative graph, and the ability to alleviate the dimensional collapse issue of item embeddings. In the future, we will attempt to incorporate item features to construct a feature-based global graph and fuse it with the refined structural graph studied in this paper.

\begin{acks}
This research was supported by grants from the National Natural Science Foundation of China (No. 62202443), and partially supported by the Young Scientists Fund of the Natural Science Foundation of Sichuan Province (No.2023NSFSC1402). Besides, we thank MindSpore\cite{MindSpore} for the support of this work.
\end{acks}

\bibliographystyle{ACM-Reference-Format}
\balance
\bibliography{references}


\end{document}